\documentclass[preprint,amsmath,amssymb,11pt,english,aps,showpacs,showkeys]{revtex4}
\usepackage{graphicx}
\usepackage{graphics}
\usepackage{amsmath}
\usepackage{dcolumn}
\usepackage{amssymb}
\usepackage{bm}
\usepackage[latin1]{inputenc}

\begin{document}
\title{Torsion and noninertial effects on a nonrelativistic Dirac particle}
\author{K. Bakke}
\email{kbakke@fisica.ufpb.br}
\affiliation{Departamento de F\'isica, Universidade Federal da Para\'iba, Caixa Postal 5008, 58051-970, Jo\~ao Pessoa, PB, Brazil.}

\begin{abstract}
We investigate torsion and noninertial effects on a spin-$1/2$ quantum particle in the nonrelativistic limit of the Dirac equation. We consider the cosmic dislocation spacetime as a background and show that a rotating system of reference can be used out to distances which depend on the parameter related to the torsion of the defect. Therefore, we analyse torsion effects on the spectrum of energy of a nonrelativistic Dirac particle confined to a hard-wall potential in a Fermi-Walker reference frame. 

\end{abstract}


\maketitle

\section{Introduction}

In recent decades, a great deal of works has studied the influence of torsion on several physical systems \cite{t1,t3,t6,t8,t9,shap,kleinert,moraesG,moraesG2,moraesG3,put}. The interaction between fermions and torsion and possible physical effects have been discussed in Refs. \cite{t8,shap,t6}. In crystalline solids, torsion has been studied in the continuum picture of defects by using the differential geometry in order to describe the strain and the stress induced by the defect in an elastic medium \cite{tt1,tt4,kat,tt7,afur,bm}. The study of the influence of torsion on quantum systems has been extended to the electronic properties of graphene sheets \cite{voz1}, Berry's phase \cite{tt10}, quantum scattering \cite{tt11}, Landau levels for a nonrelativistic scalar particle \cite{tt13} and holonomic quantum computation \cite{bf22}. The influence of torsion on a two-dimensional quantum ring has been discussed in Ref. \cite{ani} and on a quantum dot in Ref. \cite{tt8}.

In this paper, we discuss torsion effects on the spectrum of energy of a nonrelativistic spin-$1/2$ particle confined to a hard-wall confining potential in a noninertial reference frame by considering the cosmic dislocation spacetime as a background and by showing that a rotating system of reference can be used out to distances which depend on the parameter related to the torsion of the defect. Studies of noninertial effects have discovered quantum effects related to geometric phases \cite{sag,sag5,r3} and the coupling between the angular momentum and the angular velocity of the rotating frame \cite{r1,r2,r4}. In other quantum systems, the study of noninertial effects have been extended to the weak field approximation \cite{r6}, the influence of Lorentz transformations \cite{r5}, scalar fields \cite{r8}, Dirac fields \cite{r10}, persistent currents in quantum rings \cite{r11}, spin currents \cite{r12} and rotational and gravitational effects in quantum interference \cite{r14,r15,r16}. An interesting discussion made in Ref. \cite{landau3} is by making a coordinate transformation (in the Minkowski spacetime) from a system at rest to a uniformly rotating frame, then, the line element of the Minkowski spacetime is not well-defined at large distances. This means that the coordinate system becomes singular at large distances, which is associated with the velocity of the particle would be greater than the velocity of the light. In recent years, the behaviour of external fields and the wave function of a neutral particle have been analysed under the influence of noninertial effects and the presence or absence of curvature \cite{b}. In Refs. \cite{b4,b10}, it has been shown that noninertial effects can provide a confinement of a neutral particle interacting with external fields analogous to a two-dimensional quantum dot. Thereby, our focus is to analyse torsion and noninertial effects and fill a lack in the study of the quantum dynamics of a nonrelativistic Dirac particle.

This paper is structured as follows: in section II, we study the nonrelativistic quantum dynamics of a spin-$1/2$ particle in the Fermi-Walker reference frame and in the presence of torsion. Then, we show that a rotating system of reference can be used out to distances which depend on the parameter related to the torsion of the defect and use the coordinate singularity to impose where the wave function must vanish, which allows us to obtain bound states analogous to the confinement of a spin-$1/2$ quantum particle to a hard-wall confining potential; in section III, we present our conclusions.

\section{Torsion effects on the nonrelativistic quantum dynamics of a Dirac particle in a noninertial frame}

In this section, we study the confinement of a nonrelativistic Dirac particle to a hard-wall confining potential under the influence of noninertial effects and torsion. We present the mathematical tools to describe spinors under the influence of torsion and noninertial effects, when we consider the local reference frame of the observers is a Fermi-Walker reference frame. In this paper, we work with the units $\hbar=c=1$. We start by writing the line element of the cosmic dislocation spacetime \cite{moraesG2,put,dis} (in the rest frame of the observers):
\begin{eqnarray}
ds^{2}=-d\mathcal{T}^{2}+d\mathcal{R}^{2}+\mathcal{R}^{2}d\Phi^{2}+\left(d\mathcal{Z}+\zeta\,d\Phi\right)^{2}.
\label{1.1}
\end{eqnarray}
The parameter $\zeta$ is a constant, and it is related to the torsion of the defect. From the crystallography language, the parameter $\zeta$ is related to the Burgers vector $\vec{b}=b\,\hat{z}$, where $\zeta=\frac{b}{2\pi}$ \cite{kleinert,moraesG2,dis}.

In the following, we consider a coordinate transformation given by $\mathcal{T}=t$, $\mathcal{R}=\rho$, $\Phi=\varphi+\omega\,t$, and $\mathcal{Z}=z$, where $\omega$ is the constant angular velocity of the rotating frame. Thus, the line element (\ref{1.1}) becomes
\begin{eqnarray}
ds^{2}&=&-\left(1-\omega^{2}\rho^{2}-\zeta^{2}\omega^{2}\right)\,dt^{2}+\left(2\omega\rho^{2}+2\zeta^{2}\omega\right)d\varphi\,dt+d\rho^{2}+\rho^{2}d\varphi^{2}\nonumber\\
&+&2\zeta\omega\,dz\,dt+\left(dz+\zeta\,d\varphi\right)^{2}.
\label{1.5}
\end{eqnarray}
We can note that the line element (\ref{1.5}) is defined for values of the radial coordinate inside the range: 
\begin{eqnarray}
0\,<\rho\,<\frac{\sqrt{1-\zeta^{2}\omega^{2}}}{\omega}.
\label{1.5a}
\end{eqnarray}

Therefore, for all values of the radial coordinate given by $\rho\,\geq\,\frac{\sqrt{1-\zeta^{2}\omega^{2}}}{\omega}$, the line element (\ref{1.5}) is not well-defined. Observe that for values of the radial coordinate given by $\rho\,\geq\,\frac{\sqrt{1-\zeta^{2}\omega^{2}}}{\omega}$, we have that the component $g_{tt}$ of the metric tensor becomes positive. The meaning of this restriction of the radial coordinate is that, in the rotating system, the coordinate system becomes singular at $\rho\rightarrow\frac{\sqrt{1-\zeta^{2}\omega^{2}}}{\omega}$, which is associated with the velocity of the particle would be greater than the velocity of the light as discussed in Ref. \cite{landau3}. Moreover, the range (\ref{1.5a}) shows that the restriction of the radial coordinate in the rotating frame depends on the angular velocity and the parameter associated with the torsion of the defect. Therefore, based on the coordinate singularity given in Eq. (\ref{1.5a}), we analyse the behaviour of the quantum particle under the influence of torsion and noninertial effects inside the region $0\,<\,\rho<\sqrt{1-\zeta^2\omega^2}/\omega$ by imposing that the wave function vanishes at $\rho\rightarrow\frac{\sqrt{1-\zeta^{2}\omega^{2}}}{\omega}$, which yields bound states analogous to the confinement of a spin-$1/2$ particle to a hard-wall confining potential \cite{dot,dot2,dot3,ring3}.

From the line element (\ref{1.5}), we can see that we have chosen to work with curvilinear coordinates, thus, the Dirac spinors should be worked with the mathematical formulation of the spinor theory in curved space \cite{weinberg}. In a curved spacetime background, spinors are defined locally, where each spinor transforms according to the infinitesimal Lorentz transformations, that is, $\psi'\left(x\right)=D\left(\Lambda\left(x\right)\right)\,\psi\left(x\right)$, where $D\left(\Lambda\left(x\right)\right)$ corresponds to the spinor representation of the infinitesimal Lorentz group and $\Lambda\left(x\right)$ corresponds to the local Lorentz transformations \cite{weinberg}. Locally, the reference frame of the observers can be build via a noncoordinate basis $\hat{\theta}^{a}=e^{a}_{\,\,\,\mu}\left(x\right)\,dx^{\mu}$, where the components $e^{a}_{\,\,\,\mu}\left(x\right)$ are called \textit{tetrads} and satisfy the relation: $g_{\mu\nu}\left(x\right)=e^{a}_{\,\,\,\mu}\left(x\right)\,e^{b}_{\,\,\,\nu}\left(x\right)\,\eta_{ab}$ \cite{weinberg,bd,naka}, where $\eta_{ab}=\mathrm{diag}(- + + +)$ is the Minkowski tensor. The inverse of the tetrads are defined as $dx^{\mu}=e^{\mu}_{\,\,\,a}\left(x\right)\,\hat{\theta}^{a}$, and the relations $e^{a}_{\,\,\,\mu}\left(x\right)\,e^{\mu}_{\,\,\,b}\left(x\right)=\delta^{a}_{\,\,\,b}$ and $e^{\mu}_{\,\,\,a}\left(x\right)\,e^{a}_{\,\,\,\nu}\left(x\right)=\delta^{\mu}_{\,\,\,\nu}$ are satisfied. In the Fermi-Walker reference frame, noninertial effects can be observed from the action of external forces without any influence of arbitrary rotations of the spatial axis of the local frame \cite{misner}. This reference frame can be built by taking $\hat{\theta}^{0}=e^{0}_{\,\,\,t}\left(x\right)\,dt$, which means that the components of the noncoordinate basis form a rest frame for the observers at each instant, and the spatial components of the noncoordinate basis $\hat{\theta}^{i}$ must be chosen in such a way that they do not rotate \cite{misner}. With these conditions, we can write
\begin{eqnarray}
\hat{\theta}^{0}=dt;\,\,\,\,\hat{\theta}^{1}=d\rho;\,\,\,\,\hat{\theta}^{2}=\rho\,\omega\,dt+\rho\,d\varphi;\,\,\,\,\hat{\theta}^{3}=\zeta\,\omega\,dt+\zeta\,d\varphi+dz.
\label{1.8}
\end{eqnarray}

Recently, the Fermi-Walker reference frame has been used in studies of the analogue effect of the Aharonov-Casher effect \cite{bf12}, and the Dirac oscillator \cite{b9}. Hence, by solving the Maurer-Cartan structure equations in the presence of torsion $T^{a}=d\hat{\theta}^{a}+\omega^{a}_{\,\,\,b}\wedge\hat{\theta}^{b}$ \cite{naka}, where $T^{a}=\frac{1}{2}\,T^{a}_{\,\,\,\mu\nu}\,dx^{\mu}\wedge\,dx^{\nu}$ is the torsion 2-form, $\omega^{a}_{\,\,\,b}=\omega_{\mu\,\,\,\,b}^{\,\,\,a}\left(x\right)\,dx^{\mu}$ is the connection 1-form, the operator $d$ corresponds to the exterior derivative and the symbol $\wedge$ means the wedge product, we obtain the following non-null components of the connection 1-form:
\begin{eqnarray}
T^{3}&=&2\pi\zeta\,\delta\left(\rho\right)\,\delta\left(\varphi\right)\,d\rho\wedge d\varphi;\nonumber\\
\omega_{\varphi\,\,\,2}^{\,\,\,1}\left(x\right)&=&-\omega_{\varphi\,\,\,1}^{\,\,\,2}\left(x\right)=-1\,;\label{1.10}\\
\omega_{t\,\,\,\,2}^{\,\,1}\left(x\right)&=&-\omega_{t\,\,\,1}^{\,\,\,2}\left(x\right)=-\omega.\nonumber
\end{eqnarray}

In order to write the Dirac equation in a curved spacetime background and in the presence of torsion, we need to take into account that the partial derivative becomes the covariant derivative, where the covariant derivative is given by $\partial_{\mu}\rightarrow\nabla_{\mu}=\partial_{\mu}+\Gamma_{\mu}\left(x\right)+K_{\mu}\left(x\right)$, with $\Gamma_{\mu}\left(x\right)=\frac{i}{4}\,\omega_{\mu ab}\left(x\right)\,\Sigma^{ab}$ being the spinorial connection \cite{bd,naka}, and $K_{\mu}\left(x\right)=\frac{i}{4}K_{\mu ab}\left(x\right)\Sigma^{ab}$, with $\Sigma^{ab}=\frac{i}{2}\left[\gamma^{a},\gamma^{b}\right]$. The indices $(a,b,c=0,1,2,3)$ indicate the local reference frame. The connection $1$-form $K_{\mu ab}\left(x\right)$ is related to the contortion tensor by \cite{shap}: $K_{\mu ab}= K_{\beta\nu\mu}\left[e^{\nu}_{\,\,\,a}\left(x\right)\,e^{\beta}_{\,\,\,b}\left(x\right)-e^{\nu}_{\,\,\,b}\left(x\right)\,e^{\beta}_{\,\,\,a}\left(x\right)\right]$. Following the definitions of Ref. \cite{shap}, the contortion tensor is related to the torsion tensor via $K^{\beta}_{\,\,\,\nu\mu}=\frac{1}{2}\left(T^{\beta}_{\,\,\,\nu\mu}-T_{\nu\,\,\,\,\mu}^{\,\,\,\beta}-T^{\,\,\,\beta}_{\mu\,\,\,\,\nu}\right)$, where we have that the torsion tensor is antisymmetric in the last two indices, while the contortion tensor is antisymmetric in the first two indices. Moreover, it is usually convenient to write the torsion tensor into three irreducible components: the trace vector $\bar{T}_{\mu}=T^{\beta}_{\,\,\,\mu\beta}$, the axial vector $S^{\alpha}=\epsilon^{\alpha\beta\nu\mu}\,T_{\beta\nu\mu}$ and in the tensor $q_{\beta\nu\mu}$, which satisfies the conditions $q^{\beta}_{\,\,\mu\beta}=0$ and $\epsilon^{\alpha\beta\nu\mu}\,q_{\beta\nu\mu}=0$. Thus, the torsion tensor becomes: $T_{\beta\nu\mu}=\frac{1}{3}\left(\bar{T}_{\nu}\,g_{\beta\mu}-\bar{T}_{\mu}\,g_{\beta\nu}\right)-\frac{1}{6}\,\epsilon_{\beta\nu\mu\gamma}\,S^{\gamma}+q_{\beta\nu\mu}$. The $\gamma^{a}$ matrices are defined in the local reference frame and correspond to the Dirac matrices in the Minkowski spacetime \cite{bd,greiner}, \textit{i.e.},
\begin{eqnarray}
\vec{\Sigma}=\left(
\begin{array}{cc}
\vec{\sigma} & 0 \\
0 & \vec{\sigma} \\	
\end{array}\right);\,\,\,\,\,\,
\gamma^{0}=\hat{\beta}=\left(
\begin{array}{cc}
I & 0 \\
0 & -I \\
\end{array}\right);\,\,\,\,\,\,
\gamma^{i}=\hat{\beta}\,\hat{\alpha}^{i}=\left(
\begin{array}{cc}
 0 & \sigma^{i} \\
-\sigma^{i} & 0 \\
\end{array}\right),
\label{2.3}
\end{eqnarray}
with $\vec{\Sigma}$ and $I$ being the spin vector and the $2\times2$ identity matrix, respectively. The matrices $\vec{\sigma}$ are the Pauli matrices and satisfy the relation $\left(\sigma^{i}\,\sigma^{j}+\sigma^{j}\,\sigma^{i}\right)=2\eta^{ij}$. The $\gamma^{\mu}$ matrices are related to the $\gamma^{a}$ matrices via $\gamma^{\mu}=e^{\mu}_{\,\,\,a}\left(x\right)\gamma^{a}$ \cite{bd}. In this way, the general expression for the Dirac equation describing torsion effects on a quantum particle in the Fermi-Walker reference frame (\ref{1.8}) is
\begin{eqnarray}
m\psi&=&i\gamma^{0}\frac{\partial\psi}{\partial t}-i\omega\,\gamma^{0}\,\frac{\partial\psi}{\partial\varphi}+i\gamma^{1}\,\frac{\partial\psi}{\partial\rho}+i\frac{\gamma^{2}}{\rho}\left[\frac{\partial}{\partial\varphi}-\zeta\frac{\partial}{\partial z}\right]\,\psi+i\gamma^{3}\frac{\partial\psi}{\partial z}+i\gamma^{\mu}\,\Gamma_{\mu}\left(x\right)\,\psi\nonumber\\
[-2mm]\label{1.10a}\\[-2mm]
&+&\frac{1}{8}\,S^{0}\gamma^{0}\gamma^{5}\,\psi-\frac{1}{8}\,\vec{\Sigma}\cdot\vec{S}\,\psi.\nonumber
\end{eqnarray}

From the results obtained in Eq. (\ref{1.10}), we can calculate the non-null components of the spinorial connection $\Gamma_{\mu}\left(x\right)$, and obtain $i\gamma^{\mu}\,\Gamma_{\mu}=i\frac{\gamma^{1}}{2\rho}$ \cite{b4}. Moreover, from the definition of the axial vector $S^{\alpha}$ given above and the results given in Eq. (\ref{1.10}), we have that the only non-null component of the axial vector is $S^{0}=-\frac{4\pi\zeta}{\rho}\,\delta\left(\rho\right)\,\delta\left(\varphi\right)$ \cite{bf2}. Hence, for $\rho\neq0$, the Dirac equation (\ref{1.10a}) becomes
\begin{eqnarray}
i\frac{\partial\psi}{\partial t}=m\hat{\beta}\psi+i\omega\frac{\partial\psi}{\partial\varphi}-i\hat{\alpha}^{1}\left(\frac{\partial}{\partial\rho}+\frac{1}{2\rho}\right)\psi-i\frac{\hat{\alpha}^{2}}{\rho}\left(\frac{\partial}{\partial\varphi}-\zeta\frac{\partial}{\partial z}\right)\psi-i\hat{\alpha}^{3}\frac{\partial\psi}{\partial z}.
\label{1.11}
\end{eqnarray}

Now, let us discuss the nonrelativistic behaviour of the spin-$1/2$ particle. We can obtain the nonrelativistic dynamics of the spin-$1/2$ particle by writing the solution of the Dirac equation (\ref{1.11}) in the form 
\begin{eqnarray}
\psi=e^{-imt}\,\left(
\begin{array}{c}
\phi\\
\chi\\	
\end{array}\right),
\label{1.12}
\end{eqnarray} 
where $\phi$ and $\chi$ are two-spinors, and we consider $\phi$ being the ``large'' component and $\chi$ being the ``small'' component \cite{greiner}. Substituting (\ref{1.12}) into the Dirac equation (\ref{1.11}), we obtain two coupled equations of $\phi$ and $\chi$. The first coupled equation is
\begin{eqnarray}
i\frac{\partial\phi}{\partial t}-i\omega\frac{\partial\phi}{\partial\varphi}=\left[-i\,\sigma^{1}\frac{\partial}{\partial\rho}-\frac{i\,\sigma^{1}}{2\rho}-\frac{i\,\sigma^{2}}{\rho}\left(\frac{\partial}{\partial\varphi}-\zeta\frac{\partial}{\partial z}\right)-i\,\sigma^{3}\frac{\partial}{\partial z}\right]\chi,
\label{1.13}
\end{eqnarray}
while the second coupled equation is
\begin{eqnarray}
i\frac{\partial\chi}{\partial t}+2m\chi-i\omega\frac{\partial\chi}{\partial\varphi}=\left[-i\,\sigma^{1}\frac{\partial}{\partial\rho}-\frac{i\,\sigma^{1}}{2\rho}-\frac{i\,\sigma^{2}}{\rho}\left(\frac{\partial}{\partial\varphi}-\zeta\frac{\partial}{\partial z}\right)-i\,\sigma^{3}\frac{\partial}{\partial z}\right]\phi.
\label{1.14}
\end{eqnarray}
With $\chi$ being the ``small" component of the wave function, we can consider $\left|2m\chi\right|\gg\left|i\frac{\partial\chi}{\partial t}\right|$, and $\left|2m\chi\right|\gg\left|i\omega\frac{\partial\chi}{\partial\varphi}\right|$, thus, we can write 
\begin{eqnarray}
\chi\approx\frac{1}{2m}\left[-i\,\sigma^{1}\frac{\partial}{\partial\rho}-\frac{i\,\sigma^{1}}{2\rho}-\frac{i\,\sigma^{2}}{\rho}\left(\frac{\partial}{\partial\varphi}-\zeta\frac{\partial}{\partial z}\right)-i\,\sigma^{3}\frac{\partial}{\partial z}\right]\phi.
\label{1.15}
\end{eqnarray}

Substituting $\chi$ of the expression (\ref{1.15}) into (\ref{1.13}), we obtain a second order differential equation given by
\begin{eqnarray}
i\frac{\partial\phi}{\partial t}&=&-\frac{1}{2m}\left[\frac{\partial^{2}}{\partial\rho^{2}}+\frac{1}{\rho}\frac{\partial}{\partial\rho}+\frac{1}{\rho^{2}}\left(\frac{\partial}{\partial\varphi}-\zeta\frac{\partial}{\partial z}\right)^{2}+\frac{\partial^{2}}{\partial z^{2}}\right]\phi+\frac{i}{2m}\frac{\sigma^{3}}{\rho^{2}}\left(\frac{\partial}{\partial\varphi}-\zeta\frac{\partial}{\partial z}\right)\phi\nonumber\\
[-2mm]\label{1.16}\\[-2mm]
&+&\frac{1}{8m\rho^{2}}\,\phi+i\omega\frac{\partial\phi}{\partial\varphi},\nonumber
\end{eqnarray}
which corresponds to the Schr\"odinger-Pauli equation for a spin-$1/2$ particle under the influence of the noninertial effects of the Fermi-Walker reference frame, and a screw dislocation (torsion). Note that $\phi$ is an eigenfunction of $\sigma^{3}$ in Eq. (\ref{1.16}), whose eigenvalues are $s=\pm1$. Thus, we can write $\sigma^{3}\phi_{s}=\pm\phi_{s}=s\phi_{s}$. We can see that the operators $\hat{p}_{z}=-i\partial_{z}$ and $\hat{J}_{z}=-i\partial_{\varphi}$ \cite{schu} commute with the Hamiltonian of the right-hand side of (\ref{1.16}), thus, we write the solution of (\ref{1.16}) in terms of the eigenvalues of the operator $\hat{p}_{z}=-i\partial_{z}$, and the $z$-component of the total angular momentum $\hat{J}_{z}=-i\partial_{\varphi}$ \footnote{It has been shown in Ref. \cite{schu} that the $z$-component of the total angular momentum in cylindrical coordinates is given by $\hat{J}_{z}=-i\partial_{\varphi}$, where the eigenvalues are $\mu=l\pm\frac{1}{2}$.}: 
\begin{eqnarray}
\phi_{s}=e^{-i\mathcal{E}t}\,e^{i\left(l+\frac{1}{2}\right)\varphi}\,e^{ikz}\,\left(
\begin{array}{c}
R_{+}\left(\rho\right)\\
R_{-}\left(\rho\right)\\	
\end{array}\right),
\label{1.17}
\end{eqnarray}
where $l=0,\pm1,\pm2,\ldots$ and $k$ is a constant. Substituting the solution (\ref{1.17}) into the Schr\"odinger-Pauli equation (\ref{1.16}), we obtain two noncoupled equations for $R_{+}$ and $R_{-}$. After some calculations, we can write the noncoupled equations for $R_{+}$ and $R_{-}$ in the following compact form: 
\begin{eqnarray}
R_{s}''+\frac{1}{\rho}R_{s}'-\frac{\nu_{s}^{2}}{\rho^{2}}R_{s}+\eta^{2}\,R_{s}=0,
\label{1.18}
\end{eqnarray}
where we have defined the following parameters:
\begin{eqnarray}
\nu_{\pm}&=&\nu_{s}=l+\frac{1}{2}\left(1-s\right)-\zeta\,k\nonumber\\
[-2mm]\label{1.19}\\[-2mm]
\eta^{2}&=&2m\left[\mathcal{E}+\omega\left(l+\frac{1}{2}\right)-\frac{k^{2}}{2m}\right].\nonumber
\end{eqnarray}

The second order differential equation (\ref{1.18}) corresponds to the Bessel differential equation \cite{abra}. The general solution of (\ref{1.18}) is given by: $R_{s}\left(\rho\right)=A\,J_{\nu_{s}}\left(\eta\rho\right)+B\,N_{\nu_{s}}\left(\eta\rho\right)$, where $J_{\nu_{s}}\left(\eta\rho\right)$ and $N_{\nu_{s}}\left(\eta\rho\right)$ are the Bessel function of first and second kinds \cite{abra}. In order to have a regular solution at the origin, we must take $B=0$ in the general solution of Eq. (\ref{1.18}), since the Neumann function diverges at the origin. Thus, the solution of (\ref{1.18}) becomes: $R_{s}\left(\rho\right)=A\,J_{\left|\nu_{s}\right|}\left(\eta\rho\right)$. Moreover, we wish to obtain a normalized wave function inside the region $0\,<\,\rho\,<\,\frac{\sqrt{1-\omega^{2}\zeta^{2}}}{\omega}$, therefore we consider the spin-$1/2$ quantum particle is confined to a hard-wall confining potential by imposing that the radial wave function vanishes at $\rho\rightarrow\rho_{0}=\frac{\sqrt{1-\omega^{2}\zeta^{2}}}{\omega}$, that is,
\begin{eqnarray}
R_{s}\left(\rho\rightarrow\rho_{0}=\frac{\sqrt{1-\omega^{2}\zeta^{2}}}{\omega}\right)=0.
\label{1.20}
\end{eqnarray}

Then, by assuming that $\eta\rho_{0}\gg1$ (where $\rho_{0}=\frac{\sqrt{1-\omega^{2}\zeta^{2}}}{\omega}$), and we can take \cite{abra,b10}
\begin{eqnarray}
J_{\left|\nu_{s}\right|}\left(\eta\rho_{0}\right)\rightarrow\sqrt{\frac{2}{\pi\eta\rho_{0}}}\,\cos\left(\eta\rho_{0}-\frac{\left|\nu_{s}\right|\pi}{2}-\frac{\pi}{4}\right).
\label{1.21}
\end{eqnarray}

Hence, substituting (\ref{1.21}) into (\ref{1.20}), we obtain
\begin{eqnarray}
\mathcal{E}_{n,\,l}\approx\frac{1}{2m}\frac{\omega^{2}}{\left(1-\omega^{2}\zeta^{2}\right)}\left[n\,\pi+\frac{\pi}{2}\,\left|l+\frac{1}{2}\left(1-s\right)-\zeta\,k\right|+\frac{3\pi}{4}\right]^{2}+\frac{k^{2}}{2m}-\omega\left[l+1/2\right].
\label{1.22}
\end{eqnarray}

Equation (\ref{1.22}) is the spectrum of energy of a nonrelativistic Dirac particle confined to a hard-wall confining potential under the influence of torsion and noninertial effects. We can observe the influence of torsion on the energy levels (\ref{1.22}) in the effective angular momentum given by $\nu_{s}=l+\frac{1}{2}\left(1-s\right)-\zeta\,k$, and by the presence of the fixed radius $\rho_{0}=\frac{\sqrt{1-\omega^{2}\zeta^{2}}}{\omega}$. Note that the spectrum of energy obtained in Eq. (\ref{1.22}) is proportional to $n^{2}$ (parabolic energy spectrum) in contrast to recent studies of the influence of noninertial effects on the Landau quantization for neutral particles \cite{b}, whose spectrum of energy is proportional to the quantum number $n$. This results from the imposition of having the wave function being normalized in the space restricted by the range (\ref{1.5a}), where we have considered the wave function vanishing at $\rho\rightarrow\rho_{0}=\frac{\sqrt{1-\omega^{2}\zeta^{2}}}{\omega}$ and assumed $\eta\rho_{0}\gg1$. It is worth mentioning an analogy between the present study and studies of confinement of quantum particle to a quantum dot in condensed matter systems. The result obtained in Eq. (\ref{1.22}) agrees with the studies of the confinement of quantum particles to a quantum dot made in Refs. \cite{dot,dot2,dot3,ring3} where the quantum dot models provide a parabolic spectrum of energy (in relation to the quantum number $n$). The main difference between the present work and the quantum dot models of Refs. \cite{dot,dot2,dot3,ring3} is that the hard-wall confining potential is determined by the geometry of the manifold (described by the line element (\ref{1.5})). Hence, the geometrical approach yielded in the present work can be useful in studies of noninertial effects in condensed matter systems possessing the presence of screw dislocations \cite{moraesG2,b10}. 

We also obtain in Eq. (\ref{1.22}) the Page-Werner \textit{et al.} term \cite{r1,r2,r4}, which corresponds to the coupling between the angular velocity $\omega$ and the quantum number $l$. Returning to the analogy with condensed matter systems, we have that the presence of the Page-Werner \textit{et al.} term in Eq. (\ref{1.22}) agrees with the studies of the confinement of a neutral particle to a quantum dot analogous to the Tan-Inkson model \cite{tan} made in Ref. \cite{b4}. Finally, we can see in the last term of Eq. (\ref{1.22}) that there is no influence of the torsion on the Page-Werner \textit{et al.} term \cite{r1,r2,r4}.

\section{conclusions}

In this brief report, we have discussed torsion and noninertial effects on the confinement of a nonrelativistic Dirac particle to a hard-wall confining potential. We have seen, in the rotating system, that the coordinate system becomes singular at $\rho\rightarrow\frac{\sqrt{1-\omega^{2}\zeta^{2}}}{\omega}$, where the restriction of the values of the radial coordinate depends on the parameter related to the torsion of the defect and the angular velocity. We have also shown that by considering the wave function vanishing at $\rho\rightarrow\rho_{0}=\frac{\sqrt{1-\omega^{2}\zeta^{2}}}{\omega}$ and $\eta\rho_{0}\gg1$, we can obtain bound states whose spectrum of energy is parabolic in relation to the quantum number $n$. We also have shown that the influence of torsion on the energy levels is given by the presence of an effective angular momentum $\nu_{s}=l+\frac{1}{2}\left(1-s\right)-\zeta\,k$ and by a fixed radius $\rho_{0}=\frac{\sqrt{1-\omega^{2}\zeta^{2}}}{\omega}$. We have also obtained the coupling between the angular velocity $\omega$ and the quantum number $l$, which is known as the Page-Werner \textit{et al.} term \cite{r1,r2,r4}. Moreover, this study has shown that there exists no influence of the torsion on the Page-Werner \textit{et al.} term \cite{r1,r2,r4}. 

We would like to add a comment on the influence of torsion and noninertial effects in quantum systems. It has been shown in Ref. \cite{moraesG3} that the presence of torsion modifies the electromagnetic field in the rest frame of the observers. Following the study made in Ref. \cite{moraesG3}, it should be interesting to consider either a charged particle or a neutral particle interacting with external fields in the presence of torsion in a noninertial reference frame. Both torsion and noninertial effects can yield new field configurations and, consequently, new contributions to the energy levels can be obtained. Furthermore, the geometrical approach used in this work can be useful in studies of quantum dots in condensed matter systems described by the Dirac equation such as graphene \cite{voz1,graf}, topological insulators \cite{dirac2} and cold atoms \cite{dirac3}. A different context of using the present geometrical approach can be on the behaviour of the Dirac oscillator \cite{b9} in the cosmic dislocation spacetime and the Casimir effect \cite{mota,mb1}. Another interesting quantum effect arises from the presence of a quantum flux in the energy levels of bound states called persistent currents. For instance, in Ref. \cite{ring}, persistent currents arise from the dependence of the energy levels on the Berry phase \cite{berry} and the Aharonov-Anandan quantum phase \cite{anan}. In Ref. \cite{r11}, persistent currents have been investigated in a quantum ring from rotating effects. Hence, the study of persistent currents should be interesting in a scenario where there exists the presence of torsion in an elastic medium in a noninertial frame.

\acknowledgments

The author would like to thank CNPq (Conselho Nacional de Desenvolvimento Cient\'ifico e Tecnol\'ogico - Brazil) for financial support.

\end{document}